\documentclass[twocolumn,           
               showpacs,            
               preprintnumbers,     
               aps,                 
               prd,          	    
               a4paper,             
               superscriptaddress,  
               unsortedaddress,     
               nofootinbib,         
               tightenlines,        
               floats               
               ]{revtex4}

\usepackage{graphicx}
\usepackage{latexsym}
\usepackage{amsmath,amssymb}        
\usepackage[draft=false]{hyperref}

\let\include\input

\newcommand{\Mnodim}{\left(\frac{M}{M_4}\right)}

\newcommand{\lnodim}{\left(\frac{l}{l_4}\right)}

\newcommand{\tLnodim}{\left(\frac{t_{{\rm evap}}}{t_4}\right)}

\newcommand{\gcosm}{g_{{\rm cosm}}}
\newcommand{\tevap}{t_{{\rm evap}}}
\newcommand{\TBH}{T_{{\rm BH}}}
\newcommand{\be}{\begin{equation}}
\newcommand{\ee}{\end{equation}}
\newcommand{\bea}{\begin{eqnarray}}
\newcommand{\eea}{\end{eqnarray}}
\renewcommand{\L}{\left}
\newcommand{\R}{\right}

\begin{document}

\title{Primordial black holes in braneworld cosmologies:\\ Formation, 
cosmological evolution and evaporation}
\author{Raf Guedens}
\affiliation{DAMTP, Centre for Mathematical Sciences,
 Cambridge University, Wilberforce Road, 
            Cambridge CB3 0WA, United 
Kingdom}
\author{Dominic Clancy}
\affiliation{Astronomy Centre, University of Sussex, 
             Brighton BN1 9QJ, United 
Kingdom}
\author{Andrew R.~Liddle}
\affiliation{Astronomy Centre, University of Sussex, 
             Brighton BN1 9QJ, United 
Kingdom}
\date{\today} 
\pacs{98.80.Cq \hfill astro-ph/0205149}
\preprint{astro-ph/0205149}

\begin{abstract}
We consider the population evolution and evaporation of primordial black holes 
in the simplest braneworld cosmology, Randall--Sundrum type~II. We demonstrate 
that black holes forming during the high-energy phase of this theory (where the 
expansion rate is proportional to the density) have a modified evaporation law, 
resulting in a longer lifetime and lower temperature at evaporation, while those 
forming in the standard regime behave essentially as in the standard cosmology. 
For sufficiently large values of the AdS radius, the high-energy regime can be 
the one relevant for primordial black holes evaporating at key epochs such as
nucleosynthesis and the present. We examine the 
formation epochs of such black holes, and delimit the parameter regimes where 
the standard scenario is significantly modified.
\end{abstract}

\maketitle
\section{Introduction}

The idea that our observable Universe may be a brane embedded in a 
higher-dimensional bulk is one which has deep ramifications for cosmology, and 
which in particular may rewrite many of our ideas as to how the Universe evolved 
during its earliest stages. One probe of these early stages is the 
possible formation of a population of primordial black 
holes~\cite{PBH}, 
and for the standard cosmology considerable attention has been directed at 
establishing constraints both from evaporation products and from a possible 
contribution to the present dark matter density 
\cite{PBHcons,Carr75,Carr85}. 
The constraints on the formation rate are 
typically extremely strong, as after formation there is a long epoch during 
which the black hole energy density grows relative to radiation, so that even a 
modest initial fractional density can have a large impact at later stages.

Such constraints may be modified in many ways within the braneworld context. 
Thus far, the problem has only be studied in detail for the case of
large compact extra 
dimensions~\cite{ADM98,Kanti}; however in the first reference it was presumed 
that most of the radiation would be lost to the extra dimensions, whereas it is 
now believed that the emitted radiation is mostly confined to the 
brane~\cite{EHM20}. In this paper we adopt a different scenario, namely 
the simplest of the Randall--Sundrum models~\cite{RSII}, known as Type II 
(henceforth RS-II), 
where a positive-tension brane is embedded in a bulk with a negative 
cosmological constant. We will not specifically address black hole formation 
mechanisms, but seek to determine the properties and population evolution of the 
black holes after formation, setting up a framework enabling formation 
mechanisms to be tested against observational data.

There are many modifications to the standard constraints that need to be taken 
into account. At high energies there is a modified form of the Friedmann 
equation, which alters the cosmological temperature--time relation in the early 
stages as 
well as modifying the horizon mass. The temperature of a black hole of a 
given mass may be modified by the presence of the extra dimension, so that the 
masses of black holes persevering to key epochs such as nucleosynthesis and the 
present change, and the character of their final emission products is altered. 
The purpose of this paper is to determine how the key primordial black hole 
properties are modified in the simplest braneworld scenario.
In a forthcoming companion paper (Clancy {\it et al.}), we analyze the 
astrophysical 
constraints on the primordial black hole population taking into account these 
modifications.

\section{Braneworld cosmology}

\label{sec:cosm}

In the cosmological model as outlined in Ref.~\cite{branecos}, our universe is 
a positive tension brane embedded in an (otherwise empty) AdS bulk, which is 
$Z_2$ symmetric about the brane. The energy--momentum tensor of fields confined 
to the brane will be taken to be of perfect fluid form. If the metric on the 
brane is of the Friedmann--Robertson--Walker form, the Einstein equations 
projected onto the brane reduce to 
the usual energy conservation equation
\begin{equation} 
\dot \rho+3H(\rho+p)=0\,, 
\end{equation}
and a modified Friedmann equation
\begin{equation} 
H^2=\frac{8 \pi}{3 M_4^2}\left( \rho+\frac{\rho^2}{2 
\lambda}+\rho_{{\rm KK}}\right)+\frac{\Lambda_4}{3}-\frac{k}{a^2}\,.
\end{equation}
Here an overdot denotes derivative with respect to cosmic time $t$, $\rho$ and
$p$ are the energy density and pressure of the fluid, $a$ is
the scale factor on the Friedmann brane, with $H$ the Hubble constant,
$k=-1,0,1$ for open, flat or closed Friedmann branes respectively, 
$M_4$ is the effective 4D Planck mass and $\Lambda_4$ is the 4D cosmological 
constant. Furthermore, $\rho_{{\rm KK}}$ is an effective energy density 
stemming from the bulk Weyl tensor; it behaves like (dark) radiation, 
$\rho_{{\rm KK}}=\rho_{{\rm KK},0}\left( a_0/a\right)^4$, although
$\rho_{{\rm KK},0}$ needn't be positive. Finally, the brane tension
$\lambda$ is related to the fundamental 5D Planck mass $M_5$ by 
$\lambda=3M_5^6/4 \pi M_4^2$.

Defining the AdS curvature radius $l$ in terms of the bulk
cosmological constant
\begin{equation}
\Lambda_5=-\frac{3}{4\pi}\frac{M_5^3}{l^2}\,,
\end{equation}
we have 
\begin{equation}
\Lambda_4=3\left(\frac{M_5^6}{M_4^4}-\frac{1}{l^2}\right)\,.
\end{equation}
In the following, $\Lambda_4$ will be set to zero. The AdS radius provides an 
effective size of the extra dimension. As will become apparent, differences 
between RS-II and the standard scenario will be most pronounced for 
black holes whose radius is much smaller than the AdS radius.
With $\Lambda_4\equiv 0$, it follows that the brane-tension $\lambda$ and 
the AdS radius $l$ are related via 
\begin{equation}
\lambda^{-1/4}=
\left(\frac{4 \pi}{3}\right)^{1/4} \left(\frac{l}{l_4}\right)^{1/2} l_4 \,,
\label{laml}
\end{equation}
where $l_4=M_4^{-1}$ is the 4D Planck length.

In Ref.~\cite{RSII}, corrections to the Newtonian potential of a point mass
$m$ due to the 5th dimension were calculated for large distances as 
\begin{equation}
V(r)=\frac{2m}{M_4^2\; r}\left(1+\frac{2}{3}\frac{l^2}{r^2}
\right)\,.
\end{equation}
Current experiments using torsion pendulums have failed to observe such 
corrections on scales down to $r 
\approx 0.2\; \mathrm{mm}$ \cite{ref9}. This means 
the AdS radius must be smaller than $l_{{\rm max}} \equiv 
10^{31} l_4$. [To our knowledge this is the strongest upper bound on the
AdS radius to date. A much weaker constraint derives from the fact
that the high-energy phase (defined below) should be over at the onset
of nucleosynthesis, giving $l<10^{43} l_4$.]

The case of interest for primordial black hole formation is the early
universe, and we will focus on a flat radiation-dominated model. 
As for the dark radiation term, nucleosynthesis constrains
$(\rho_{{\rm KK}}/\rho)_{\mathrm{nuc}}$ to be smaller than $0.024$
\cite{ref3}. Since both energy terms scale in the same way, the dark
radiation will always have a small effect on the overall dynamics, and
will be neglected in the remainder. With these assumptions, the
solutions for the scale factor and energy density are
\begin{equation}
\rho=\frac{3 M_4^2}{32 \pi} \frac{1}{t(t+t_c)}\,,
\end{equation}
and
\begin{equation}
a=a_0 \left[\frac{t\left(t+t_c\right)}{t_0\left(t_0+t_c\right)}
\right]^{1/4}\,,
\end{equation}
where $t_0$ is any non-zero time, and $t_c$ is the `transition time' 
\begin{equation}
t_c \equiv \frac{l}{2}\,.\label{tc}
\end{equation}
At times much smaller than $t_c$ (equivalent to $\rho \gg \lambda$),
this gives rise to a non-conventional {\it high-energy} regime, in 
which
\begin{eqnarray}
a & = & a_0 \left(\frac{t}{t_0}\right)^{1/4}\,,\label{ahigh} \\
\rho & = & \frac{3 M_4^2}{32 \pi \,t_c\,t}\,,\label{rhoHE}\\
R_{{\rm H}}=4 t \,, & \quad & \quad M_{{\rm H}}=8 M_4^2\; 
\frac{t^2}{t_c}\,,\label{mhhigh}
\end{eqnarray}
with $R_{{\rm H}}$ and $M_{{\rm H}}$ denoting the Hubble radius 
and mass inside the Hubble horizon respectively.
For times much larger than $t_c$, we recover the regime of 
standard cosmology 
where
\begin{eqnarray}
a & = & a_0 \left(\frac{t}{t_0^{1/2} t_c^{1/2}}\right)^{1/2}\,, \label{astand}\\
\rho & = & \frac{3 M_4^2}{32 \pi\, t^2}\,,\\
R_{{\rm H}}=2 t \,, & \quad & M_{{\rm H}}= M_4^2 t \,.\label{mhstand}
\end{eqnarray}

In the high-energy regime, as in standard cosmology, the radiation has
a temperature given by 
\begin{equation}
\rho=\frac{\pi^2}{30}\, g_{{\rm cosm}}\; T^4\,,
\end{equation}
where $g_{{\rm cosm}}$ indicates the number of relativistic particle species 
at a particular time. This gives rise to a modified temperature--time relation:
\begin{equation}
\frac{T}{T_4}=\left(\frac{45}{8 \pi^3}\right)^{1/4} 
\gcosm^{-1/4} \lnodim^{-1/4} \left(\frac{t}{t_4}\right)^{-1/4} 
\,.\label{temptimeHE}
\end{equation}
An interesting background temperature to consider is at the transition
time between the high-energy and standard regime. Taking 
$g_{{\rm cosm}}=O(100)$, it reads
\begin{equation}
T_c=3 \times 10^{18} \lnodim^{-1/2} {{\rm GeV}}\,. \label{Tc} \end{equation}
Its minimum value allowed by experiment is $T_c(l_{{\rm max}})\approx
10^3$ GeV.

Inflation is an important part of the standard cosmology and we will assume 
that 
black hole formation takes place after it, possibly though not necessarily 
induced by inflation-generated density perturbations. There is a firm upper 
limit on the inflationary energy scale from the requirement that the 
gravitational waves it produces don't lead to excessive large-angle microwave 
anisotropies, and this leads to a lower limit on the horizon 
mass.\footnote{Inclusion of density perturbations strengthens this constraint 
somewhat, as does allowing for reduction in energy density during the late 
stages of inflation, so our limits are conservative.} The amplitude of 
gravitational waves in the RS-II model was computed in Ref.~\cite{LMW}; using 
their notation it is
\begin{equation}
A_{{\rm T}}^2 = \frac{4}{25\pi} \, \frac{H^2}{M_4^2} \, F^2(Hl) \,,
\end{equation}
where
\begin{equation}
F(x)  = \left[\sqrt{1+x^2} - x^2 \sinh^{-1} \frac{1}{x} \right]^{-1/2} \,.
\end{equation}
In the high-energy regime $x \gg1$, $F^2(x) \simeq 3x/2$.

\begin{figure}[t]
\includegraphics[width=\linewidth]{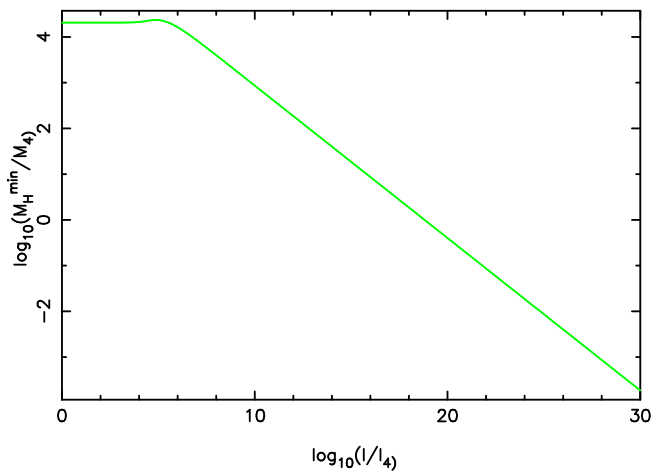}\\
\caption[horlimit]{\label{horlimit} The minimum horizon mass after inflation as 
a function of the AdS radius. For $l/l_4 \lesssim 10^5$ the constraint 
corresponds to inflation ending in the low-energy regime, whereas for larger $l$ 
it corresponds to the high-energy regime.}
\end{figure}

If we require that gravitational waves contribute no more than half the 
anisotropy signal seen by COBE (in order to leave room for density perturbations 
to induce structure formation), this gives the limit $A_{{\rm T}}^2 < 3 \times 
10^{-11}$ \cite{LL}. Combining this with the horizon mass 
formula Eq.~(\ref{mhhigh}) gives a lower limit on the horizon mass, and hence 
on the masses of PBHs that can form. In the high-energy regime it gives 
\begin{equation}
M_{{\rm H}} > 2 \times 10^6 \lnodim^{-1/3} M_4=  2 \times 10^6\; M_5\,.
\end{equation}
The general expression for the lower limit on $M_{{\rm H}}$ is shown in 
Figure~\ref{horlimit}. The limit is quite weak, with allowed initial
masses even 
below $M_4$ in the high-energy limit (though not of course below $M_5$). This 
limit does not restrict any of the situations we will consider, as black holes 
surviving to nucleosynthesis always have masses higher than this limit. One 
could however in principle have inflation models where the energy scale after 
inflation was low enough to prevent the formation of early evaporating PBHs.

\section{Black Hole Evaporation}
\label{sec:evap}

\subsection{The Evaporation Rate and Lifetime}

In this section, using standard black hole thermodynamical arguments, a 
mass--lifetime relation will be derived for black holes that formed by a small 
amount of matter collapsing on the brane.\footnote{The study of collapse 
in this context has been carried out by a number of 
authors~\cite{DadhichGhosh2001,Brunietal2001}, although 
at the time of writing a full description is lacking. 
These studies have revealed that the nature of collapse is much 
richer and more complex in the braneworld context and in 
Refs.~\cite{Brunietal2001,Dadhichetal2000} it was conjectured that primordial 
black 
holes could in principle have formed from the collapse of 
dark radiation alone. Here, however, we shall assume a minimal picture of 
collapsing matter on the brane.} 
We will also determine the
range of values of the AdS radius for which the derivation is
valid. This will be used in the final section to estimate the time of
formation of primordial black holes in the present cosmological scenario.

Consider a black hole formed from collapsing matter confined
to the brane. It will have an event horizon that extends into the
bulk. Moreover, if the size of the hole $r_0$ is much smaller than the AdS
radius $l$ (and neglecting possible charges or rotation), it is natural to 
assume its geometry is given by a 5D
Schwarzschild solution\footnote{This would certainly be the outcome 
according to a higher-dimensional generalization of the hoop 
conjecture~\cite{ADM98,Thorne}. Near the horizon, the black hole would have no 
way of distinguishing the AdS dimension from the others. See
also Ref.~\cite{hoop}.}~\cite{EHM99}
\begin{equation}
ds_{5}^2=-f(r)\, dt^2+f^{-1}(r) \, dr^2+r^2 \, d\Omega_3^2 \,, 
\label{5dmetric}\end{equation}
with $f(r)=1-r_0^2/r^2$ and $d\Omega_3$ the volume element of
a 3-sphere. This form of the metric is a good approximation in
the vicinity of the event horizon, which is the region
needed to analyze the Hawking effect. The black hole is expected
to emit Hawking radiation both into the brane and the bulk by exciting the 
brane or bulk degrees of freedom. In the present model, only gravitational 
radiation can propagate in the bulk. It is worth noting that near the horizon,
the induced metric on the brane is given by
\begin{equation}
ds_{4}^2=-f(r) \, dt^2+f^{-1}(r) \, dr^2+r^2 \, d\Omega_2^2 \,, 
\label{induced}\end{equation}
which is not the usual $4$D Schwarzschild metric.\footnote{It is
expected \cite{EHM99} that the metric will approach the standard $4$D form far 
away from the event horizon. An interesting class of exact solutions to the 
RS-II
4D brane equations which describe Reissner--Nordstr\"om type black 
holes, but possessing a so-called `tidal charge' which arises due 
to the presence of a non-zero bulk Weyl tensor, have been given by 
Dadhich {\it et al.}~\cite{Dadhichetal2000}. However, 
it is not yet clear whether these solutions are consistent with a 
full 5D solution. If so then we expect that these should represent 
a class of large black holes, i.e.~black holes formed in the 
low-energy regime.} 
Indeed, this
metric has an effective negative energy--momentum tensor outside the horizon 
that will modify the gray-body factors of radiation by quantum fields
confined to the brane \cite{EHM20}. However, the effective potentials
in the field propagation equations bear similarity to those of the
standard treatment. They also vanish when approaching the horizon,
reducing the propagation equations to free wave equations. Since the
brane is tuned to be flat, the derivation of the 
Hawking process on the brane will essentially be identical to the
standard case. As for the bulk, Hawking radiation in AdS space has
been discussed in Ref.~\cite{Hemming}, where it was shown to be similar to
the asymptotically-flat case. 

The expressions for radius, area and temperature of the black hole are given 
in terms of the AdS radius $l$ and the black hole mass $M$ as 
\begin{eqnarray}
r_0 & = &  \sqrt{\frac{8}{3 \pi}} \frac{M^{1/2}}{M_5^{3/2}} = \sqrt{\frac{8}{3 
\pi}} \lnodim^{1/2} \Mnodim^{1/2} l_4 \,, \quad \quad
\label{bhrad} \\
A_5 & = & 2 \pi^2 r_0^3 \,,\\
\TBH & = & \frac{1}{2 \pi r_0}\,, \label{tbh}
\end{eqnarray}
and hold provided $r_0 \ll l$. This is to be contrasted with the usual 
$4$D result
\begin{equation}
\label{4Dtemp}
\TBH(4{\rm D}) = \frac{M_4^2}{8\pi M} \,.
\end{equation}

To estimate the lifetime, consider the number of particles of a
certain species $j$, emitted in $D$-dimensional spacetime by a black
hole of temperature $\TBH$, in a time interval $dt$ and with momentum
in the interval $(\mathbf{k},\mathbf{k}+d\mathbf{k})$:
\begin{equation}
dN_j=\sigma_j(k)\; \frac{dt}{{\rm exp}(\omega/\TBH)\pm 1}
\;\frac{d^{D-1}k}{(2 \pi)^{D-1}}\,,\label{spec}
\end{equation}
with $\omega^2=k^2+m^2$ and $m$ the mass of the particle. The upper and lower 
sign apply to fermions and bosons respectively. As regards the 
absorption/emission 
cross-sections $\sigma_j$, summation over all angular modes
is understood. In general, they depend on the species and
frequency and must be determined numerically~\cite{Page}. 
Due to the different metrics Eq.~(\ref{induced}) and Eq.~(\ref{5dmetric}), 
accurate determination of the cross-sections is beyond the scope of this paper. 
However, in the high-frequency limit all cross-sections reduce to the
same expression (see below). In the low-frequency
limit, the cross-sections decrease with frequency, approaching a finite
value for spin-$0$ or spin-$1/2$ particles, whereas they vanish with
increasing powers of frequency for higher-spin particles. This means
the total energy emitted in higher-spin particles is suppressed
relative to particles of lowest spin. These features are expected to
carry over to the brane context, while the numerical factors may
change somewhat.
     
The rate of energy loss by $D$-dimensional evaporation is obtained
from Eq.~(\ref{spec}) as
\begin{equation}
\frac{dM}{dt}=-\;\sum_j \int \sigma_j(k)\; 
\frac{\omega}{{\rm exp}(\omega/\TBH)\pm 1}\;\frac{d^{D-1}k}{(2 \pi)^{D-1}}\,. 
\label{power}\end{equation}
In the high-frequency limit ($\omega \gg T_{BH}$) all cross-sections
become identical, namely
\begin{equation}\sigma\approx
\frac{A_{\mathrm{eff},D}}{4}\equiv 
\frac{\Omega_{D-2}\;r_{\mathrm{eff},D}^{D-2}}{4}\,,\end{equation}
where $\Omega_{D-2}$ is the volume of a $(D-2)$-sphere and
$r_{\mathrm{eff},D}$ an effective radius for black-body emission,
defined as ~\cite{EHM20,ref6} 
\begin{equation}
r_{{\rm eff},D}=\left(\frac{D-1}{2}\right)^{1/(D-3)}
\left(\frac{D-1}{D-3}\right)^{1/2} r_0 \,. 
\end{equation}
Adopting this approximation for all cross-sections reduces Eq.~(\ref{power})
to Stefan's law: 
\begin{equation}
\frac{dM}{dt}\approx -g_{D}\;\tilde \sigma_D
\,A_{\mathrm{eff},D}\, T^D, \end{equation}
with $g_D$ composed of bosonic and fermionic degrees of freedom as
\begin{equation}g_D=g_{D,\mathrm{bos}}+\frac{2^{D-1}-1}{2^{D-1}}\,g_{D,\mathrm{f
erm}}
\,.\end{equation}
Further, $\tilde \sigma_D$ denotes the $D$-dimensional
Stefan--Boltzmann constant, defined per degree of freedom:
\begin{equation}\tilde \sigma_D= \frac{\Omega_{D-2}}{4\;(2 \pi)^{D-1}}\;
\Gamma(D)\, \zeta(D),\end{equation}
with $\zeta(D)$ the Riemann zeta function.\footnote{The $D$-dimensional
Stefan--Boltzmann constant was misrepresented in Ref.~\cite{EHM20}.} 

\begin{figure}[t]
\includegraphics[width=\linewidth]{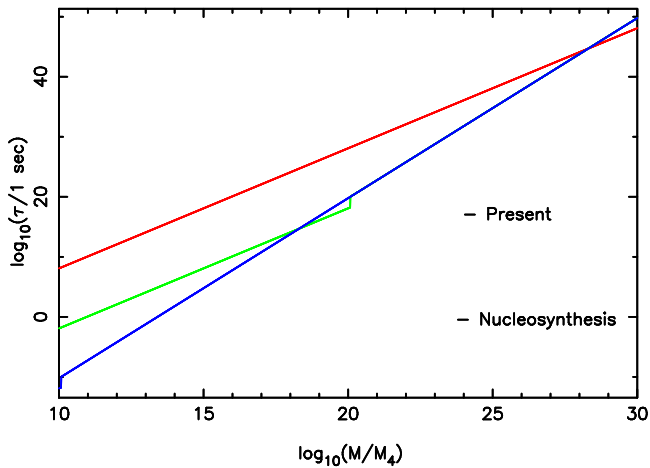}\\
\caption[lifetime]{\label{lifetime} The lifetime of black holes of different 
initial masses, for the choices $l/l_4=10^{10}$, $10^{20}$ and $10^{30}$ (from 
bottom to top), approximating $\tilde{g} = 0.032$ for all masses. For the lowest 
$l$ the usual 4D result applies across all the mass range, and for intermediate 
$l$ the 
discontinuity arises from the mismatch of the 4D and 5D relations across the 
transition. Lifetimes corresponding to nucleosynthesis and to the present age of 
the Universe are indicated.}
\end{figure}

In the present set-up, we thus estimate the total emitted power as
\begin{equation}
\frac{dM}{dt}\approx -g_{\mathrm{brane}}\tilde \sigma_4
\tilde{A}_{\mathrm{eff},4} T^4-g_{\mathrm{bulk}}\tilde \sigma_5 
A_{\mathrm{eff},5}T^5 \,,
\label{rate}\end{equation}
where we must take $\tilde{A}_{\mathrm{eff},4} = 4\pi r_{\mathrm{eff},5}^2$ 
because of the induced metric Eq.~(\ref{induced}).
Further, $g_{{\rm brane}}$ and $g_{{\rm bulk}}$ denote the brane and bulk 
degrees
of freedom with rest masses lower than $\TBH$.
Since we have regarded $g_{{\rm bulk}}$ from the 5D point of view, it does not 
count the number of graviton Kaluza--Klein modes. Rather, it is the number of 
polarization states of the graviton, $D(D-3)/2$, which gives 
$g_{{\rm bulk}}=5$. 
The number of quantum fields into which the hole
evaporates is approximately constant until its lifetime is nearly
over. Then substituting the relevant expressions into
Eq.~(\ref{rate}) and integrating gives the lifetime $\tevap$ of a 
black hole of initial mass $M$: 
\begin{equation}
\frac{\tevap(M)}{t_4}\approx \tilde g^{-1} \frac{l}{l_4} \,
\left(\frac{M}{M_4}\right)^2 \,,\label{life}
\end{equation}
with
\begin{equation}
\tilde g \equiv \frac{1}{160} \; g_{\mathrm{brane}}+
\frac{9 \zeta(5)}{32 \pi^4} \; g_{\mathrm{bulk}} \,.
\end{equation}
In the standard BH thermodynamics, Stefan's law was shown to overestimate the 
emitted power Eq.~(\ref{power})
by a factor $2.6$~\cite{Page}, and we therefore divide the
first term of $\tilde g$ by the same factor, which should remain approximately 
true. The overestimate should be at least as severe for the bulk gravitational 
radiation, because of the higher spin suppression and the confining
influence of the negative cosmological constant (although the latter
supposedly would have little influence on small black holes); a more thorough
analysis is required to be definite, but we divide by the same factor 2.6, 
resulting in a corrected form
\begin{equation}
\tilde g\approx 0.0024\;g_{{\rm brane}} +0.0012\;g_{{\rm bulk}} \,.
\end{equation}
Since only gravity is allowed to propagate in the bulk, $g_{\rm bulk}$ 
simply counts the number of polarization states of the graviton, 
namely $D(D-3)/2=5$. Combined with the fact that $g_{{\rm brane}}>g_{{\rm 
bulk}}$, it is now apparent that evaporation into the bulk
is a subdominant effect, even for very small black holes. 
We mention two typical values for $\tilde g$: If the black hole emits only 
massless particles we have $g_{\mathrm{brane}}=7.25$ and $\tilde
g\approx 0.023$. If the hole is just hot enough to emit electron--positron 
pairs, we have  $g_{\mathrm{brane}}=10.75$ and $\tilde g\approx
0.032$. Given the fairly qualitative nature of current observational 
constraints, results will be rather insensitive to the
precise value of $\tilde g$.

By comparing with the lifetime of a black hole of same mass in standard 
relativity 
\begin{equation}
\frac{\tevap(M,4{\rm D})}{t_4} \approx
1.2\times 10^4\;g_{\mathrm{brane}}^{-1}\, \L(\frac{M}{M_4}\R)^3 \,,
\label{4dlifetime}
\end{equation}
we find 
\begin{equation}
\frac{\tevap(M,5{\rm D})}{\tevap(M,4{\rm D})} \sim \L(\frac{l}{r_0(5
{\rm D})}\R)^2\,.
\end{equation}
For a fixed mass, small black holes can have much longer lifetimes in the 
higher-dimensional case. 

Figure~\ref{lifetime} shows the lifetimes of black holes for three choices of 
the AdS radius. As will become clear in the following sections, for 
$l=10^{20} l_4$, corresponding to a brane tension 
$\lambda^{1/4} = 10^9 {\rm GeV}$, black holes initially of 
the AdS radius would be evaporating around the present epoch, and so this $l$ 
marks the transition between whether presently-evaporating black holes are 
effectively four or five dimensional. For values of $l$ higher than this, black 
holes evaporating today have lower initial masses than the usual $10^{20} \, M_4 
\simeq 10^{15} \, {\rm g}$.

\begin{figure}[t]
\includegraphics[width=\linewidth]{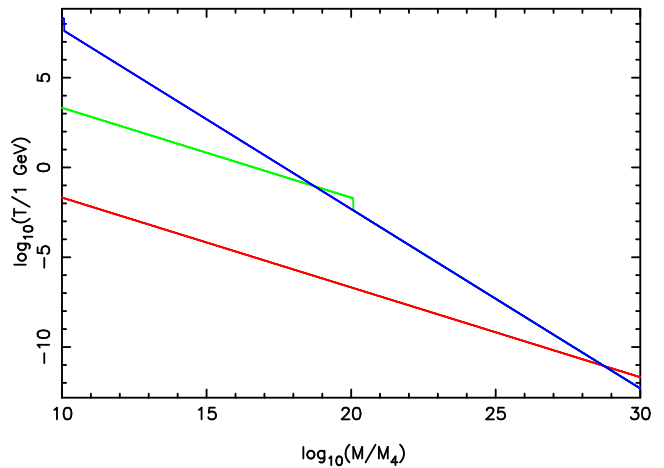}\\
\caption[temperature]{\label{temperature} As Figure~\ref{lifetime} but showing 
the initial temperature, with the top line corresponding to the lowest $l$.}
\end{figure}

Figure~\ref{temperature} shows the initial temperatures of black holes for the 
same choices of $l$. Most of the energy of a PBH is radiated at temperatures 
close to the initial temperature, with only a small fraction in a high-energy 
tail as the evaporation culminates. For $l \gtrsim 10^{20} l_4$, the 
temperature of black holes 
evaporating at the present is reduced.

For future reference, we list the $5\rm{D}$ expressions for 
mass and temperature in terms of the lifetime 
\begin{eqnarray}
\frac{M}{M_4} & = & \tilde g^{1/2}
\left(\frac{\tevap}{t_4}\right)^{1/2} \lnodim^{-1/2} \,, 
\label{masslife}\\
\frac{\TBH}{T_4} & = & \sqrt{\frac{3}{32 \pi}}\; \tilde g^{-1/4} 
\tLnodim^{-1/4} \lnodim^{-1/4}\,,\label{templife}
\end{eqnarray}
where $T_4=M_4$ is the 4D Planck temperature.

\subsection{Ranges}
\label{subsec:ranges}

The mass--lifetime relation Eq.~(\ref{masslife}) was derived under the 
assumption that the initially formed black hole is small, $r_0 \ll l$. Using
the mass--radius relation Eq.~(\ref{bhrad}), this implies the consistency
condition
\begin{equation}
\frac{M}{M_4} \ll \frac{l}{l_4} \,.
\label{smallmass} 
\end{equation}
Thus, for a black hole of a given $\it{lifetime}$ there is an allowed range of 
values of $l$ for which it is small, obtained by substituting
the mass--lifetime relation into condition Eq.~($\ref{smallmass}$): 
\begin{equation} 
l_{{\rm min}} \ll l \ll l_{{\rm max}}\,, \label{lrange} 
\end{equation}
with 
\begin{equation}
l_{{\rm min}}(\tevap) =\tilde g^{1/3} \tLnodim^{1/3} \l_4 \,, 
\label{lminmax}
\end{equation}
and $l_{{\rm max}}$ the experimental upper limit on the AdS radius quoted 
earlier. 

Using Eqs.~(\ref{masslife}) and (\ref{templife}), this corresponds to a range
on the initial mass and temperature. The mass ranges from
\begin{eqnarray}
M_{{\rm min}} & \equiv & M(\tevap,l_{{\rm max}}) \nonumber \\ 
& = & 
\left(\frac{l_{{\rm max}}}{l_4}\right)^{-1/2} \tilde g^{1/2} \tLnodim^{1/2}M_4 
\end{eqnarray}
to
\begin{equation}
M_{{\rm max}}\equiv M(\tevap,l_{{\rm min}})=\tilde g^{1/3} 
\tLnodim^{1/3}\;M_4\,.
\end{equation}
As for the black hole temperature, it ranges from
\begin{eqnarray}
T_{{\rm BH,min}} & \equiv & \TBH(\tevap,l_{{\rm max}})  \\
& = & \sqrt{\frac{3}{32 \pi}}
\left(\frac{l_{{\rm max}}}{l_4}\right)^{-1/4} \tilde g^{-1/4} \tLnodim^{-1/4}  
T_4 \nonumber
\end{eqnarray}
to
\begin{eqnarray}
T_{{\rm BH,max}} & \equiv & \TBH(\tevap,l_{{\rm min}}) \nonumber \\
& =& \sqrt{\frac{3}{32 \pi}}\; \tilde g^{-1/3} \tLnodim^{-1/3} T_4\,.
\end{eqnarray}

We note that, although the braneworld scenario allows
PBHs of a given lifetime to be lighter than in the standard case,
their initial temperature will be lower as well.

The maximum values are essentially what is obtained
in the standard 4D theory, where $M \approx 
0.04\;g_{\mathrm{brane}}^{1/3}\; (\tevap/t_4)^{1/3}\;M_4$  and 
$\TBH \approx g_{\mathrm{brane}}^{-1/3}\;
(\tevap/t_4)^{-1/3}\; T_4$. This should come as no surprise,
since they correspond to the limit of what can be considered a 
small black hole. Well beyond that limit, i.e.~for much smaller values of 
the AdS radius, the initial size (on the brane) of a black hole of the same 
lifetime would have been large, and it would have started out with properties 
indistinguishable from a $4$D black hole \cite{EHM99}. At a
certain stage, the size of the hole will become comparable with the AdS 
radius, a transition stage that has so far eluded accurate description. But 
this happens near the end of its lifetime, when most of its mass has
evaporated.\footnote{Although the lifetime in its 5D phase will be longer as
compared to the standard estimates, for a given black hole it will still be
a short time as compared to the 4D phase.}
For those black holes, we use the conventional estimates for the 
mass--lifetime relation etc. 

Two examples are of particular interest in terms of observational consequences. 
The first concerns PBHs with lifetime equal
to the present age of the universe. The currently-favoured low-density flat 
cosmology has an age of about 14 gigayears, i.e.~$t_{0}\approx 8 \times 
10^{60}\;t_4$.
The AdS radius marking the transition between 4D and 5D behaviour is 
$l_{{\rm min}}=7 \times 10^{19}\, l_4$.    
The mass then ranges from $M_{{\rm max}}=4 \times 10^{14} {\rm g}$ in the
standard scenario, to $M_{{\rm min}}=3 \times 10^9 {\rm g}$ for 
$l=l_{{\rm max}}= 10^{31}\;l_4$. The allowed temperatures range from  
$T_{{\rm BH,max}}= 25\; \mathrm{MeV}$ in the standard scenario to 
$T_{{\rm BH,min}}=50\,\mathrm{keV}$ for $l=l_{{\rm max}}$. Note that 4D 
PBHs are hot enough to copiously emit electrons, whereas 
only massless Standard Model particles can be emitted for large values of 
the AdS radius.

As a second example, consider the era of nucleosynthesis 
($t_{{\rm nuc}}\approx 100 \, {\rm s}\approx 10^{44}\,t_4$). Taking
$g_{{\rm brane}}=100$, we find $l_{{\rm min}}=10^{15}\, l_4$.
The mass ranges from $M_{{\rm max}}=5 \times 10^{9} {\rm g}$ in the
standard scenario to $M_{{\rm min}}=2 \times 10^2 {\rm g}$ for 
$l=l_{{\rm max}}$. The temperatures range from  
$T_{{\rm BH,max}}= 2 \times 10^3\, \mathrm{GeV}$ to 
$T_{{\rm BH,min}}= 0.2\, \mathrm{GeV}$.

\begin{figure}[t]
\includegraphics[width=\linewidth]{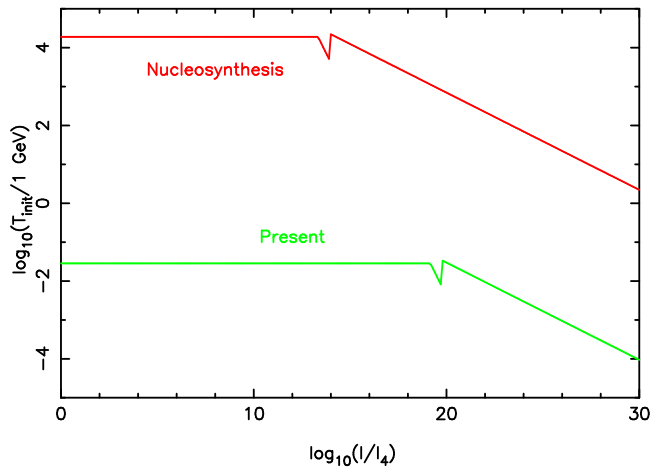}\\
\caption[inittemp]{\label{inittemp} The initial temperature of black holes 
evaporating at the key epochs of nucleosynthesis and the present, shown as a 
function of $l$.}
\end{figure}

The initial temperatures of PBHs evaporating at these two epochs are shown as a 
function of $l$ in Figure~\ref{inittemp}.

\section{Formation and Evolution}

We now return to the cosmology of Section~\ref{sec:cosm}. There are several
mechanisms by which black holes could have formed in the early
universe (see Ref.~\cite{Carr85} for a review.) We focus on collapse of 
primordial
density fluctuations.

\subsection{Formation mass}

The end stage of the collapse process is highly 
nonlinear, and it is difficult to be very precise, as the formation masses 
remain poorly understood even in the standard cosmology. However it can be 
argued that the mass of the hole will be of order the Hubble horizon mass at the 
time of formation, $t_i$, by studying the Jeans mass.
Consider a slightly overdense region of energy density $\tilde
\rho$. Its density contrast is defined as $\delta=(\tilde \rho
-\rho)/\tilde\rho$. Expanding the density contrast in Fourier modes, 
perturbation 
theory provides evolution equations for these modes, as long
as $\delta \ll 1$. The Jeans length $L_{{\rm J}}$ is then defined such that 
modes with wavelength bigger than $L_{{\rm J}}$ are growing modes, while
those with smaller wavelength oscillate. In Ref.~\cite{ref10} the 
mode equations for the Friedmann model were given for a braneworld
scenario. Applied to the present model, they read
\begin{equation}
\ddot \delta_k+H\;\dot \delta_k+\left[-\frac{16 \pi}{3
M_4^2}\;\rho-\frac{24 \pi}{M_4^2}\;\frac{\rho^2}{\lambda}+\frac{1}{3}
\left(\frac{k}{a}\right)^2\right] \delta_k=0 \,. \label{modeq}
\end{equation}
Here, $k$ denotes the comoving wavenumber of the mode. The Jeans
length is obtained by setting the expression in the brackets to zero. In the
high-energy regime we can neglect the first term, leading to
\begin{equation}
L_{{\rm J}}=\frac{\pi}{3}\sqrt{\frac{2}{3}} R_H\approx 0.85\; R_H.
\end{equation}
Just as in standard cosmology, in the high-energy regime the Jeans
length is of order the horizon size.
 
The scenario for forming PBHs is the standard one.  One starts with a slightly
overdense region in a flat FRW Universe, on a scale much larger than the
horizon.  Because of the superhorizon scale, the region can separately be
described as a portion of a closed FRW model~\cite{Harrison, Carr75}.
Therefore, it will expand less rapidly than the environment, and the density 
contrast will grow.  At a certain time the region will cross the Jeans scale.  
If the density contrast is still very small at that time, its evolution will be 
accurately described by Eq.~(\ref{modeq}), and it will start to oscillate, 
preventing collapse from ever occurring. Thus, a necessary condition for black 
hole collapse is $\delta \gtrsim 1$ at `Jeans' crossing, which as shown above 
is more or less at horizon crossing.

To keep account of the uncertainty in the precise formation mass, 
we introduce a 
factor $f$ as
follows: 
\begin{equation}
M_i=f\; M_H(t_i)\,.\label{MM}
\end{equation}
A certain amount of controversy exists over the possible 
range of $f$, although, recent numerical studies seem to
favour $f\sim 1$~\cite{NJ}. Moreover, we find in general that the
constraints examined in the companion paper will turn out not 
to be too sensitive to its exact value. 

\subsection{Formation time}

First consider PBHs forming in the high-energy regime. Then by assumption it
holds that $t_i \ll t_c$. By substituting the expressions 
($\ref{mhhigh}$) and ($\ref{tc}$) for horizon radius and transition
time, this translates into
\begin{equation} 
R_H(t_i)\ll 2\, l \,.\label{smallhor}
\end{equation}
Since the mass of the PBH is not expected to be larger than the horizon
mass, its initial radius will not be larger than the horizon.   
It is clear from Eq.~($\ref{smallhor}$) that PBHs formed in the 
high-energy regime are $\it{small}$ and effectively 
5-dimensional.
The formation time can be expressed in terms of the initial mass or
lifetime, by substituting Eqs.~(\ref{mhhigh}) and (\ref{tc}) into 
Eq.~(\ref{MM}) and using the mass--lifetime relation Eq.~(\ref{masslife}):
\begin{eqnarray}
\frac{t_i}{t_4} & =& 
\frac{1}{4}f^{-1/2}\left(\frac{M_i}{M_4}\right)^{1/2}\lnodim^{1/2} \nonumber \\
 & = & \frac{1}{4} f^{-1/2} \tilde g^{1/4} \tLnodim^{1/4} \lnodim^{1/4} \,. 
\label{tF}
\end{eqnarray}

In the standard regime, the formation time is given through
Eq.~(\ref{mhstand}) as
\begin{equation}
\frac{t_i}{t_4}=f^{-1} \frac{M_i}{M_4}=0.04\; f^{-1}
g_{\mathrm{brane}}^{1/3} \tLnodim^{1/3} \,,
\label{tFstand}
\end{equation}
which now must satisfy $t_i \gg t_c$. Thus $f^{-1} M/M_4 \gg
0.5\;l/l_4$, which violates condition Eq.~(\ref{smallmass}). As
could be expected, a black hole formed in the standard 
regime will be large, and behaves for the vast majority of its lifetime as a 
4D object.

Using Eqs.~(\ref{bhrad}) and (\ref{mhhigh}) it is straightforward to 
show that
\be
r_0 =f^{1/2} \sqrt{\frac{8}{3\pi}} 4 t_i \approx f^{1/2} R_H,
\ee
i.e. that at the formation time, $t_i$, the Schwarzschild radius 
associated with the collapsing perturbation, of mass $M_i$,
is of order the horizon size, $R_H$, so long as 
$f$ is not much smaller than 1. This is important since it implies 
that the collapsing perturbation will fall within its Schwarzschild radius 
and so form a black hole very soon after entering the horizon. Hence, 
as with the standard PBH scenario it is reasonable 
to assume that we need not concern ourselves too greatly 
with details such as the anisotropy and inhomegeneity 
of the collapse in the nonlinear subhorizon regime, since the 
black hole should form before any such effects have a chance to act.

Finally, we note that the minimum mass enforced by the condition that
PBHs form after inflation guarantees that their mass will be much greater than 
the Planck mass relevant at that time (either $M_5$ in the high-energy regime 
or $M_4$ in the low-energy regime), which in turn means that their lifetime 
will be much greater than the formation time.

\subsection{Evolution}

Once the black holes have formed, their evolution must be followed forwards in 
time to the epoch where observational constraints might apply, either the 
present epoch or the time of evaporation. As the PBH comoving number density is 
constant 
up until evaporation, as usual the relative density of PBHs as compared to 
radiation will grow proportional to the scale factor while evaporation is 
negligible. A common approximation is to presume that evaporation is negligible 
right up until the lifetime of the PBH is reached, at which point its entire 
mass--energy is released with products characteristic of its {\em initial} 
temperature, and this approximation continues to be good in the braneworld case.

\section{Conclusions}

We have carried out a detailed investigation of how primordial black hole 
scenarios are modified in the RS-II braneworld. Whether these changes are 
significant depends on the AdS radius $l$ of the braneworld model; if this is 
sufficiently small then the standard scenario is recovered. However, current 
constraints on the AdS radius are very weak ($l \lesssim 10^{31}\, l_4$ where 
$l_4$ is the 4-dimensional Planck length), and substantial modifications to the 
usual case are possible for black holes evaporating at any epoch. If the AdS 
radius exceeds $10^{15}\, l_4$ then the properties of PBHs evaporating at 
nucleosynthesis (or earlier) are modified, and if it exceeds $10^{20} l_4$ PBHs 
evaporating up to the present epoch are affected. PBHs will have modified 
evolution if and only if they form during the high-energy phase of braneworld 
cosmological evolution.

If braneworld effects are important, they act to reduce the mass of a black hole 
surviving to a given epoch. More importantly, they give a reduced temperature, 
which will alter the evaporation products characteristic of such PBHs.
An important application of these results is to investigate how constraints on 
PBH abundances are modified in the braneworld scenario. Because the 
black holes surviving to key epochs such as nucleosynthesis and the present can 
have modified temperatures, the standard astrophysical constraints cannot be 
applied and must be rederived from scratch. We carry out this analysis in a 
forthcoming companion paper (Clancy {\it et al.}).

Throughout we have ignored the possibility that PBHs might grow significantly 
through accretion of the background, known to be a valid approximation in the 
standard cosmology \cite{Carr75}. However in the high-energy regime this issue 
deserves re-investigation, which we will do in a forthcoming paper.

We have considered the simplest of braneworlds. It would be
interesting to see how robust our conclusions are in more complicated
cosmological models, for instance those including one or more bulk
fields. Unless their number is very large, this will not drastically alter the 
energy fraction a black hole loses to the
brane as compared to the bulk. On the other hand, the early phase of 4D
cosmology can be significantly modified, in turn altering the relation
between the black hole's lifetime and time of formation. This is left for 
future investigation. 


\begin{acknowledgments}
D.C.~was supported by PPARC and A.R.L.~in part by the Leverhulme Trust. 
R.G.~would like to thank John Barrow for inspiration, and we thank 
Carsten van de Bruck, Bernard Carr, Anne Green, and also Roy Maartens 
and the group at ICG Portsmouth for discussions.
\end{acknowledgments}




                
\def\jrnld#1#2#3#4#5{{#1}, {#2} {\bf #3}, {#4} ({#5}).} 

\def\jrnl#1#2#3#4#5{{#1}, {#2} {\bf #3}, {#4} ({#5})} 

\def\jrnlna#1#2#3#4{{#1} {\bf #2}, {#3} ({#4})}

\def\jrnlT#1#2#3#4#5#6{{#1}, {\em #2}, {#3}, {\bf #4}, {#5} ({#6}).}


\def\jrnlE#1#2#3#4#5#6{{#1}, {#2} {\bf #3}, {#4}  ({#5}), {\tt #6}.}

\def\jrnlTE#1#2#3#4#5#6#7{{#1}, {\em #2}, {#3}, {\bf #4}, {#5} ({#6}), {\tt 
#7}.}


\def\jrnltwo#1#2#3#4#5#6#7#8#9{ {#1}, {#2} {\bf #3}, {#4} (#5); {#6} {\bf #7}, 
{#8} {(#9)}.}

\def\preprint#1#2#3#4{{#1}, {\em #2}, {\tt #3} ({#4}).}


\def\Book#1#2#3#4{{#1}, {\em #2}, {#3} (#4).} 

\def\Bookpage#1#2#3#4#5{{#1}, {\em #2}, {#3}, {#4} ({#5}).} 

\def\edBookpage#1#2#3#4#5#6#7{{#1}, {\em #2} in {#3}, Ed. {#4}, {#5}, {#6} 
({#7}).}


\def\Proc#1#2#3#4#5#6#7{{#1}, {\em #2}, Proc. {#3 }, {Ed. #4}, {#5} ({#6}), {\tt 
#7}.}



\def\ADM{Adv. Math.}
\def\ANP{Ann. Phys. (NY)}
\def\ARM{Arch. Math.}
\def\ASA{Astron. Astrophys.}
\def\ASJ{Astrophys. J.}
\def\ASP{Astropart. Phys.}
\def\ASZ{Astron. Zh.}
\def\CMP{Commun. Math. Phys.}
\def\CQG{Class. Quant. Grav.}
\def\GRG{Gen. Rel. Grav.}
\def\IJM{Int. J. Mod. Phys.}
\def\JHE{J. High Energy Phys.}
\def\JLM{J. London Math. Soc.}
\def\JMP{J. Math. Phys.}
\def\JOP{J. Phys.}
\def\MAA{Math. Ann.}
\def\MNR{Mon. Not. Roy. Ast. Soc.}
\def\MPL{Mod. Phys. Lett.}
\def\MAZ{Math. Zeit.} 
\def\NUC{Nuovo Cimento}
\def\NIM{Nucl. Instrum. Methods}
\def\NAT{Nature}
\def\NUP{Nucl. Phys.}
\def\PHL{Phys. Lett.}
\def\PHR{Phys. Rep.}
\def\PHR{Phys. Rev.}
\def\PRL{Phys. Rev. Lett.}
\def\PRS{Proc. Roy. Soc. Lon.} 
\def\PTR{Phil. Trans. Roy. Soc. Lon.}
\def\PTP{Prog. Theor. Phys.}
\def\RAP{Rel. Astro. Phys.}
\def\RMP{Rev. Mod. Phys.}
\def\RPP{Rep. Prog. Phys.}
\def\SPJ{Sov. Phys. JETP} 
\def\SOA{Sov. Astron.}
\def\TEN{Tensor N. S.}
\def\ZEP{Z. Phys.}
\def\ZET{Zh. Eksp. Teor. Fiz.}



\def\ACP{Academic Press Inc.}
\def\ADW{Addison-Wesley Publishing Company: Redwood City, California}
\def\CUP{Cambridge University Press: Cambridge}
\def\DOV{Dover Publications, New York}
\def\ELS{Elsevier Science B. V., Amsterdam}
\def\FRE{W. H. Freeman, New York}
\def\HUP{Harvard University Press: Cambridge, Massachusetts}
\def\IOP{Institute of Physics Publishing: Bristol and Philadelphia}
\def\KLU{Kluwer: Dordrecht}
\def\MCH{McGraw-Hill Book Company: New York}
\def\PUP{Princeton University Press, Princeton: New Jersey}
\def\SVN{Springer-Verlag: New York}
\def\SVB{Springer-Verlag: Berlin-Heidelberg}
\def\UCP{University of Chicago Press: Chicago}
\def\WIL{John Wiley and Sons Ltd: Chichester}
\def\WOS{World Scientific: Singapore}



\end{document}